# Rydberg States of $H_3$ and HeH as Potential Coolants for Primordial Star Formation

*Gokul Kannan,[†] Jeremy R. Chien,[†] Anthony J. Benjamin,[†] Niranjan Bhatia,[‡] Richard J. Saykally*[†]*

†Department of Chemistry, University of California, Berkeley,
Berkeley, CA 94720, United States

‡Monta Vista High School, 21840 McClellan Rd,
Cupertino, CA 95014, United States

Email: saykally@berkeley.edu
Office Telephone: (510) 642-8269
Lab Telephone: (510) 642-1047

## Abstract

Current theory and measurements establish the age of the universe as ca. 13.8 billion years. For the first several hundred million years of its existence, it was a dark, opaque void. After that, the hydrogen atoms comprising most of the "ordinary" matter began to condense and ionize, eventually forming the first stars that would illuminate the sky. Details of how these "primordial" stars formed have been widely debated, but remain elusive. A central issue in this process is the mechanism by which the primordial gas (mainly hydrogen and helium atoms) collected via the action of dark matter cools and further accretes to fusion densities. Current models invoke collisional excitation of $H_2$ molecular rotations and subsequent radiative rotational transitions allowed by the weak molecular quadrupole moment. In this article, we review the salient considerations, and present some new ideas, bases on recent spectroscopic observations of neutral $H_3$ Rydberg electronic state emission in the mid-infrared.

## Introduction

The early stages of primordial star formation have been the subject of extensive study for decades. For these first stars to form, it has been proposed that primordial gas clouds first condensed at nodes in a filamentary network organized by dark matter. These high density nodes proceeded to condense further through radiative cooling processes until they reached the Jeans mass, viz. the critical mass for gravity to become the dominant force driving star formation. However, as these nodes compressed, they would have gained massive amounts of energy, reaching temperatures above $10^3$ K. To sustain compression, the nodes would have had to cool via some highly effective radiative mechanism. Cooling in modern descriptions of secondary star-forming regions features extensive use of metal-based mechanisms–where "metal" refers to any elements heavier than H, He, and Li–but these fail in the primordial context due to the nearly exclusive presence of hydrogen and helium (Mac Low & Klessen, 2004). Elucidating the requisite cooling mechanisms remains a major unsolved problem in the field.

Cooling mechanisms operative from $10^9$ K to $10^4$ K are well-understood. Gas is cooled primarily via bremsstrahlung radiation (the deflection of charged particles by other charged particles, converting kinetic energy to photons). This drove the formation of "minihalos"– i.e. the "nodes" mentioned previously. It has been shown theoretically that each of these minihalos likely produced one single metal-free star, probably limited to ~300 solar masses (Abel et al., 2002). The size of these first stars is related to the Jeans mass for gravitational collapse, which is proportional to the square of the temperature and inversely proportional to the square root of the pressure. Pressures in primordial gas clouds were probably similar to those in star forming clouds studied today (Widicus Weaver and Herbst, 2019), leaving temperature as the primary variable in determining the Jeans mass, and thus as the defining factor in determining first star mass. Accordingly, efficient cooling of these stars to well below $10^4$ K is essential.

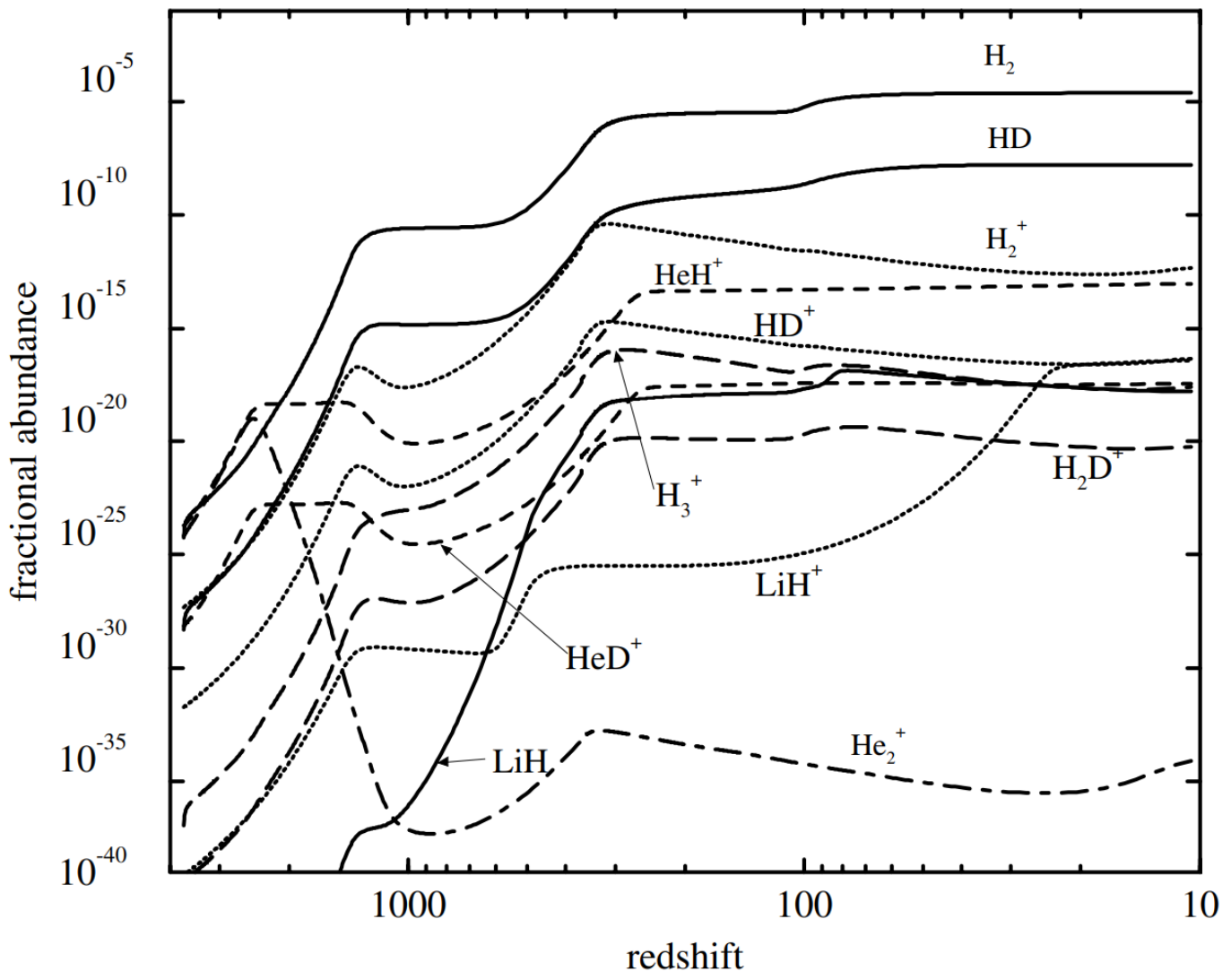

**Figure 1:** Fractional abundances of molecules and molecular ions present in the early universe at high redshift z. The cosmological parameters correspond to the Standard Model III of Stancil et al (1998): $\Omega_0 = 1$, $\Omega_b = 0.0367$, h = 0.67 (Lepp et al. 2002)

A number of simple primordial molecules have been identified as preceding star formation, and their abundances have been estimated theoretically (Fig. 1). The formation and subsequent reactions of these molecules are limited by the presence of only $^1$H, $^2$H (or D), $^3$He, $^4$He and a trace amount of $^7$Li formed in the Big Bang. Accordingly, we can list most of the possible reactions according to the standard model (Table 1). Lepp and coworkers have examined these reactions in more detail (Lepp et al., 2002). For the purposes of this paper, we are concerned primarily with the formation of $H_2$, HeH, $H_3$, and their derivatives. These molecules and their corresponding reactions give us a "toolbox" with which to begin examining potential primordial cooling mechanisms. However, we must recognize that this "toolbox" is limited; modeling abundances, reaction kinetics, and thermodynamics in primordial conditions remains an extraordinarily difficult task.

| | Reaction | $a_1$[a] | $a_2$ | $a_3$ (K) | Notes[d] |
|---|---|---|---|---|---|
| (1) | $H + e^- \rightarrow H^+ + 2e^-$ | 7.57 − 10[b] | 5.50 − 1 | 1.582 + 5[c] | 1 |
| (2) | $D + e^- \rightarrow D^+ + 2e^-$ | 7.57 − 10 | 5.50 − 1 | 1.582 + 5[c] | 2 |
| (3) | $He + e^- \rightarrow He^+ + 2e^-$ | 4.12 − 10 | 0.5 | 2.85 + 5[c] | 3 |
| (4) | $He^+ + e^- \rightarrow He^{2+} + 2e^-$ | 9.84 − 11 | 0.5 | 6.32 + 5[c] | 4 |
| (5) | $Li + e^- \rightarrow Li^+ + 2e^-$ | 3.11 − 8 | 1.63 − 1 | $6.27 + 4^{3}$ | 4 |
| (6) | $Li^+ + e^- \rightarrow Li^{2+} + 2e^-$ | 5.67 − 12 | 7.15 − 1 | 8.77 + 5[c] | 4 |
| (7) | $Li^{2+} + e^- \rightarrow Li^{3+} + 2e^-$ | 1.7 − 12 | 7.09 − 1 | 1.42 + 6[c] | 4 |
| (8) | $He^+ + e^- \rightarrow He + \nu$ | 3.66 − 7 | −1.5 | 4.7 + 5[c] | 5 |
| (9) | $He^{2+} + e^- \rightarrow He^+ + \nu$ | 1.97 − 11 | −6.34 − 1 | 1.32 + 6 | 6 |
| (10) | $Li^+ + e^- \rightarrow Li + \nu$ | 1.33 − 8 | −1.5 | 6.53 + 5[c] | 7 |
| (11) | $Li^{2+} + e^- \rightarrow Li^+ + \nu$ | 1.87 − 11 | −6.41 − 1 | 1.76 + 6 | 6 |
| (12) | $Li^{2+} + e^- \rightarrow Li^+ + \nu$ | 5.34 − 8 | −1.23 | $9.23 + 5^{3}$ | 7 |
| (13) | $Li^{3+} + e^- \rightarrow Li^{2+} + \nu$ | 4.83 − 11 | −6.21 − 1 | 1.67 + 6 | 6 |
| (14) | $H^+ + D \rightarrow H + D^+$ | 1.97 − 9 | −4.02 − 1 | 3.71 + 1 | 8 |
| | | −1.63 − 13 | 1.49 | — | |
| (15) | $D^+ + H \rightarrow D + H^+$ | 1.97 − 9 | −3.95 − 1 | 3.30 + 1 | 8 |
| | | 3.04 − 10 | −3.33 − 1 | — | |
| (16) | $H^+ + H^- \rightarrow H + H$ | 1.40 − 7 | −4.87 − 1 | −2.93 + 4 | 9 |
| (17) | $H^+ + D^- \rightarrow H + D$ | 1.61 − 7 | −4.87 − 1 | −2.93 + 4 | 10 |
| (18) | $H^- + D^+ \rightarrow H + D$ | 1.61 − 7 | −4.87 − 1 | −2.93 + 4 | 10 |
| (19) | $D^+ + D^- \rightarrow D + D$ | 1.96 − 7 | −4.87 − 1 | −2.93 + 4 | 10 |
| (20) | $Li^+ + H^- \rightarrow Li + H$ | 2.93 − 7 | −4.77 − 1 | −2.32 + 4 | 9 |
| (21) | $Li^- + H^+ \rightarrow Li + H$ | 1.80 − 7 | −4.77 − 1 | −2.32 + 4 | 7 |
| (22) | $Li^+ + D^- \rightarrow Li + H$ | 3.71 − 7 | −5.10 − 1 | −4.41 + 4 | 11 |
| (23) | $Li^- + D^+ \rightarrow Li + D$ | 2.28 − 7 | −5.10 − 1 | −4.41 + 4 | 12 |
| (24) | $H + \nu \rightarrow H^+ + e^-$ | 1.14 + 7 | 0.98 | 1.57 + 5[c] | 13 |
| (25) | $D + \nu \rightarrow D^+ + e^-$ | 1.14 + 7 | 0.98 | 1.57 + 5[c] | 2 |
| (26) | $He + \nu \rightarrow He^+ + e^-$ | 1.11 + 7 | 1.23 | 2.80 + 5[c] | 14 |
| (27) | $He^+ + \nu \rightarrow He^{2+} + e^-$ | 5.45 + 5 | 1.63 | 5.90 + 5[c] | 15 |
| (28) | $Li + \nu \rightarrow Li^+ + e^-$ | 5.08 + 6 | 1.45 | 6.05 + 4[c] | 14 |
| (29) | $Li^+ + \nu \rightarrow Li^{2+} + e^-$ | 9.23 + 6 | 1.37 | 8.46 + 5[c] | 15 |
| (30) | $Li^{2+} + \nu \rightarrow Li^{3+} + e^-$ | 5.96 + 2 | 2.63 | 1.27 + 6[c] | 13 |
| (31) | $H^- + \nu \rightarrow H + e^-$ | 2.08 + 4 | 2.13 | 8.82 + 3[c] | 14 |
| (32) | $D^- + \nu \rightarrow D + e^-$ | 2.08 + 4 | 2.13 | 8.82 + 3[c] | 2 |
| (33) | $Li^- + \nu \rightarrow Li + e^-$ | 5.29 + 5 | 1.40 | 8.10 + 3[c] | 14 |
| (34) | $H_2 + D^+ \rightarrow HD + H^+$ | 1.60 − 9 | — | — | 16 |
| (35) | $H_2^+ + H^- \rightarrow H_3^+ + e^-$ | 2.70 − 10 | −4.85 − 1 | −3.12 + 4 | 17 |

**Table 1:** Gas-phase reactions relevant to first star formation and their rate coefficients. (Lepp et. al. 2002)

[a] The rate coefficient fits presented on this table are given by the relation $\alpha = a_1(T/300)^{a_2}\exp(-T/a_3)$. (Lepp et al. 2002)

[d] Indicated reactions have an exponential term of $\exp(-a_3/T)$.

To cool from $10^4$ K to less than $10^3$ K, the currently accepted mechanism is via $H_2$ associative detachment (AD): $H + H^- \rightarrow H_2^- \rightarrow H_2^* + e^-$.

Molecular hydrogen is also widely accepted as the predominant coolant at lower temperatures in primordial gas clouds, largely due to its high abundance and cooling effectiveness at temperatures relevant to the time period of interest (< $10^4$ K). This mechanism of cooling proceeds through the AD reaction, wherein translational energy from atomic H and $H^-$ is converted into internal energy of an excited $H_2$ through rotational-vibrational excitation of the molecule upon collision. The (nonpolar) molecule can then radiatively relax via electric quadrupole transitions, and the emitted photons can escape, cooling the gas (Abel et al., 2002). Despite the apparent simplicity of this reaction, theory and experiment have failed to converge in terms of the magnitude and energy dependence of the corresponding rate coefficient (Glover et al., 2006).

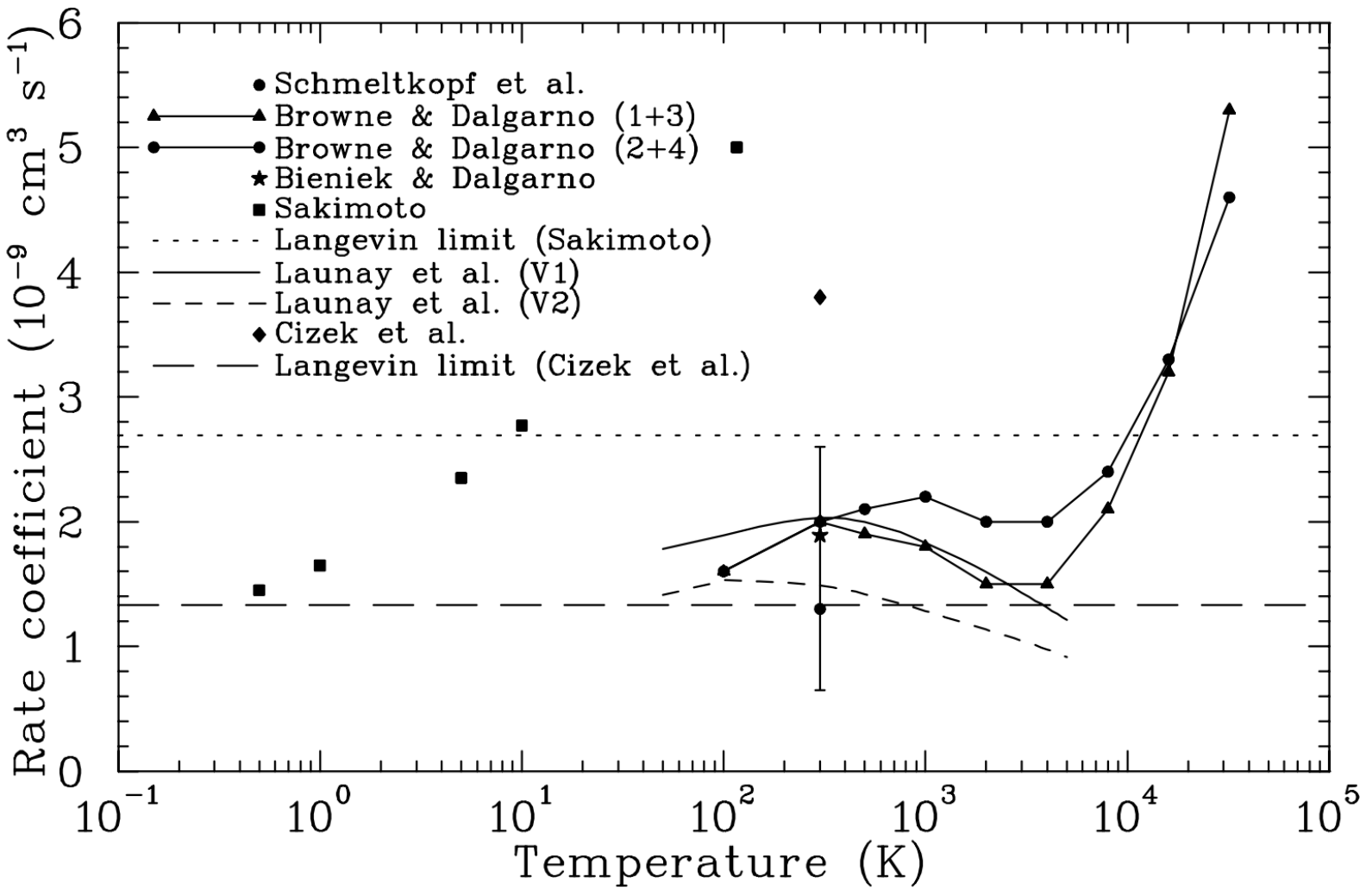

**Fig. 2:** Various published rate coefficients of $H_2$ AD at relevant early universe temperatures show large disagreements. (Glover et al., 2006)

The only experiments measuring the rate coefficient of $H_2$ AD at cosmological temperatures were carried out in 1967 by Schmeltekopf and coworkers. Virtually all early universe chemical models use their results. Subsequent computational studies of the reaction have mostly arrived at different numbers, mainly due to disagreements as to which reactions consume $H^-$ and which reactions produce negligible effects that can be omitted for simplification.

Despite the discrepancy in values of the rate coefficient, $H_2$ AD still remains the most widely accepted cooling mechanism of primordial gas clouds due to the high reactant abundance and cooling efficiency.

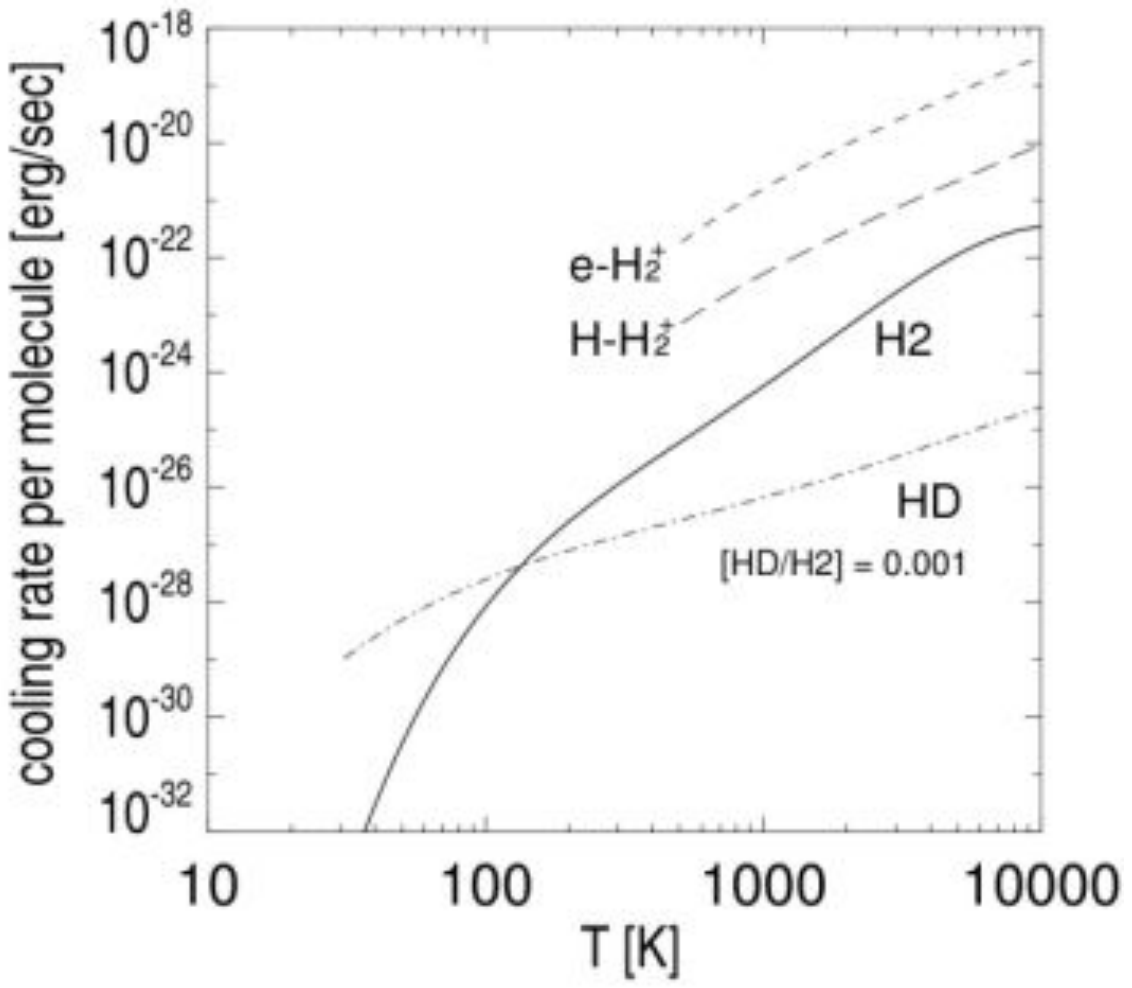

**Fig. 3**: Theoretical cooling rate per molecule vs. temperature in primordial gas environments. (Yoshida, Oh, Kitayama, & Hernquist, 2007)

This mechanism, however, falls short when considering the predictions of early star masses. $H_2$ -based cooling becomes ineffective at lower temperature, 200-300 K, as well as at sufficiently high density, establishing a lower limit for the temperature and thus for the Jeans mass, of the first stars–greater than 1000 solar masses, sufficient for supermassive stars and black holes to form instead. With the evolution of new theoretical results suggesting gravitational fragmentation of approximately one fifth of the young first-generation stars, it is possible that this mechanism is the driver in cases of fragmentation (Clark et al., 2011). However, no experimental data supporting such fragmentation have been observed (Fraser et al., 2017).

## $H_3$ Rydberg Molecules

For the minihalos to effectively cool below 200 K, a non-$H_2$ based mechanism seems necessary, as the $H_2$ electric quadrupole-allowed rotational transitions are extremely weak. The driving molecule would have to have a high formation rate at low temperatures, and necessary intermediates must have sufficiently large rate constants and long lifetimes to allow effective cooling mechanisms to proceed. We find an interesting candidate in the recombination of $H_3^+$ with a free electron, producing an excited neutral $H_3$ Rydberg molecule intermediate.

Rydberg molecules have dissociative ground states and stable non-valence excited states, parallel to those of the H atom in classic Bohr theory, and comprising orbitals energy-ordered by the increasing principal quantum number *n*. These excited species lose energy via strong dipole-allowed rovibrational radiative transitions as the electrons relax from high to lower *n* states, and they have dense manifolds of rovibrational energy levels as well (Herzberg, 1987). This efficient radiative energy loss mechanism presents an opportunity for Rydberg molecules to release large amounts of energy on relatively short timescales–a potentially effective mechanism of cooling.

Herzberg first described $H_3$ Rydberg molecule emission spectra in the 1980s and detailed the transitions between the $n = 2$ and $n = 3$ Rydberg states. While the ground state of this molecule is unbound and very short-lived, the excited (Rydberg) states are metastable (Herzberg, 1987) (Dabrowski and Herzberg, 1980) (Herzberg and Watson, 1980) (Herzberg, 1979). As such, they either decay radiatively with lifetimes proportional to $n^3$ or pre-dissociatively. As a prototype Rydberg molecule, $H_3$ has been studied extensively since Herzberg's discovery by both theory and experiment, although not until recently in the context of primordial star cooling (Saykally et al., 2010)(Wright, 2011)(Saykally, 2015).

If $H_3^+$ forms favorably in primordial environments, then in collisions with H and He atoms, rotational transitions would be excited similar to $H_2$ molecule excitation. However, these $H_3^+$ excitations would have much stronger coupling to translation due to the net charge, so subsequently formed Rydberg states would likely exhibit rovibrationally excited ion cores, in addition to the electronic energy of the molecular n-state. This would engender detectable UV-visible and IR emissions, which could ultimately radiate away translational energy.

$H_3^+$ dissociative recombination (DR) occurs through an indirect mechanism involving the neutral $H_3$ Rydberg intermediate:

$$H_3^+ + e^- \rightarrow H_3^* \rightarrow \text{neutral products}$$

The $H_3^+$ DR process has been historically difficult to study due to two major factors: First, the vibrational degrees of freedom have 3 dimensions, usually requiring accessible theoretical models to reduce dimensionality and thus to lose information. Second, there have been major discrepancies in rate coefficients, cross sections, and various other metrics obtained via different experimental designs, similar to the issues described for $H_2$ AD.

Experimental study has involved 1) storage ring experiments and 2) plasma afterglow experiments. The storage ring experiments comprise a particle accelerator, wherein energetic molecules are collided to drive reactions of interest. Plasma afterglow experiments utilize high-temperature, low-pressure plasmas, which also produce chemical reactions. While storage ring experiments yield results approximating the theoretical rate coefficients ($\sim 10^{-7}$ cm$^3$ s$^{-1}$), plasma afterglow experiments yield rates approximately an order of magnitude lower (Kokoouline et al., 2001).

Theoretical treatments of $H_3^+$ DR prior to the early 2000s had neglected the Jahn-Teller effect, a non-Bohr-Oppenheimer phenomenon accounting for energy level splitting of degenerate electronic states due to geometric distortion. However, in 2001, Kokoouline and coworkers showed that the inclusion of Jahn-Teller coupling provided dissociative recombination rates overlapping the range of experimental observations. In 2003, Kokoouline and Greene extended this theory to unify treatment of all $D_{3h}$ triatomic ions, while maintaining all mechanical degrees of freedom, with resulted in agreement with storage ring recombination rates.

In 1998, Mitchell and Johnsen described a merged-beam experiment showing a five-fold increase in the DR product signal when the product ion deflection field was reduced to 200 V cm$^{-1}$ from 3000 V cm$^{-1}$. They speculated that this was due to the field ionization of DR products, i.e. at high field strengths, the products would be re-ionized and would not be detected. $H_3$ Rydberg molecules would be the prime candidates for re-ionization, and as such, could have easily been missed in high-field experiments. Based on these data, Mitchell and Johnsen suggested that $H_3$ Rydberg molecules could have accounted for as much as 80% of the DR products.

In 2006, Glosik continued a series of promising afterglow experiments in Prague using cavity ringdown spectroscopy(Saykally et al., 2010) to record spectra of triatomic isotopomers in discharged mixtures of hydrogen and deuterium ($H_3^+$, $H_2D^+$, $HD_2^+$) originating from low rotational states. In subsequent low-temperature ion trap experiments at MPIK Heidelberg and TU Chemnitz, Glosik recorded near-infrared spectra of $H_3^+$, $H_2D^+$, and $HD_2^+$, and characterized the trapped ions at temperatures near 10 K. These low-energy interactions are directly relevant to interstellar conditions, especially in cold dense regions of space where deuterated molecules are abundant (Glosik et al., 2006) (Zymak et al., 2011).

The Prague experiments performed prior to 2006 found that DR of $H_3^+$ is pressure dependent; the DR rate of $H_3^+$ rapidly falls when $[H_2] < 10^{12}$ $cm^{-3}$. These results suggested that DR of $H_3^+$ is not exclusively a binary process and that previously measured DR rates were collective/effective (binary + ternary) values. This was the first time data suggested that ternary recombination is significant. In 2008, a flowing afterglow Langmuir probe experiment found similar pressure dependences of $D_3^+$ DR on $D_2$ pressure. A $D_2$ low density limit was measured to give a pure binary recombination rate coefficient, while a new data analysis method improved determination of effective recombination rates (Kreckel et al., 2005).

By 2010, a new Prague cryo-FALP design allowed afterglow experiments compatible with temperatures from 77 to 300 K and pressures from 400 to 2000 Pa to measure temperature and pressure dependences of $H_3^+$ recombination rates in He buffer gas (Korolev et al., 2008). The effective recombination rate increased with increasing $[H_2]$ and [He] from $\sim 10^{11}$-$10^{13}$ $cm^{-3}$ and $\sim 10^{17}$- 5 x $10^{17}$ $cm^{-3}$, respectively; however, the rate dropped significantly when $[H_2] < 10^{12}$ $cm^{-3}$ because lower concentrations cannot keep ortho- and para- $H_3^+$ in thermal equilibrium under recombination conditions (recombination is faster than re-thermalization at this concentration). The effective recombination rate increased with increasing temperature from 77 to 170 K but decreased at higher temperatures (Glosik et al., 2009) (Varju et al., 2011) (Kotrik et al., 2010). These results parallel later studies of $D_3^+$ recombination.

These authors describe $H_3^+$ ternary recombination as a two-step mechanism wherein the electron-$H_3^+$ collision creates a rotationally excited, neutral $H_3$ molecule, which then collides with He to form a "very long-lived Rydberg state with high orbital momentum." The rotationally excited $H_3$ and $D_3$ molecules respectively have lifetimes up to 300 ps and 1000 ps before autoionizing (Glosik et al., 2010). This process proved to be 100 times more effective than Thomson, Bates, and Khare's ternary recombination process (Bates and Khare, 1965). Para-$H_3^+$ was found to have a larger He-collision rate than ortho-$H_3^+$, and this difference widens below 170 K, which is responsible for small measurement and calculation discrepancies for the ternary coefficient rate at that time. Including the ternary process in calculating the effective recombination rates and buffer gas factor represented the first time results from afterglow plasma experiments agreed with ion storage ring experiments and theory.

In 2010, Saykally et al. observed unexpected infrared lasing lines while performing cavity ringdown spectroscopy experiments on a pulsed supersonic rare gas plasma of hydrogen and rare gases. Upon consulting with Professor Chris Green, these were determined to be d-p lasing lines emitted from the $H_3$ Rydberg molecule. Calculations showed that while the d-states were long-lived, the p-states relaxed quickly, ostensibly providing an efficient lasing mechanism. The experimental-theoretical comparisons are included in Figure 4. While the experimental conditions under which these laser lines were discovered are not directly relevant to primordial gas conditions, their discovery and characterization of is a significant step towards understanding the energetics of excited $H_3$ Rydberg molecules (Saykally, 2015) (Saykally et al., 2010).

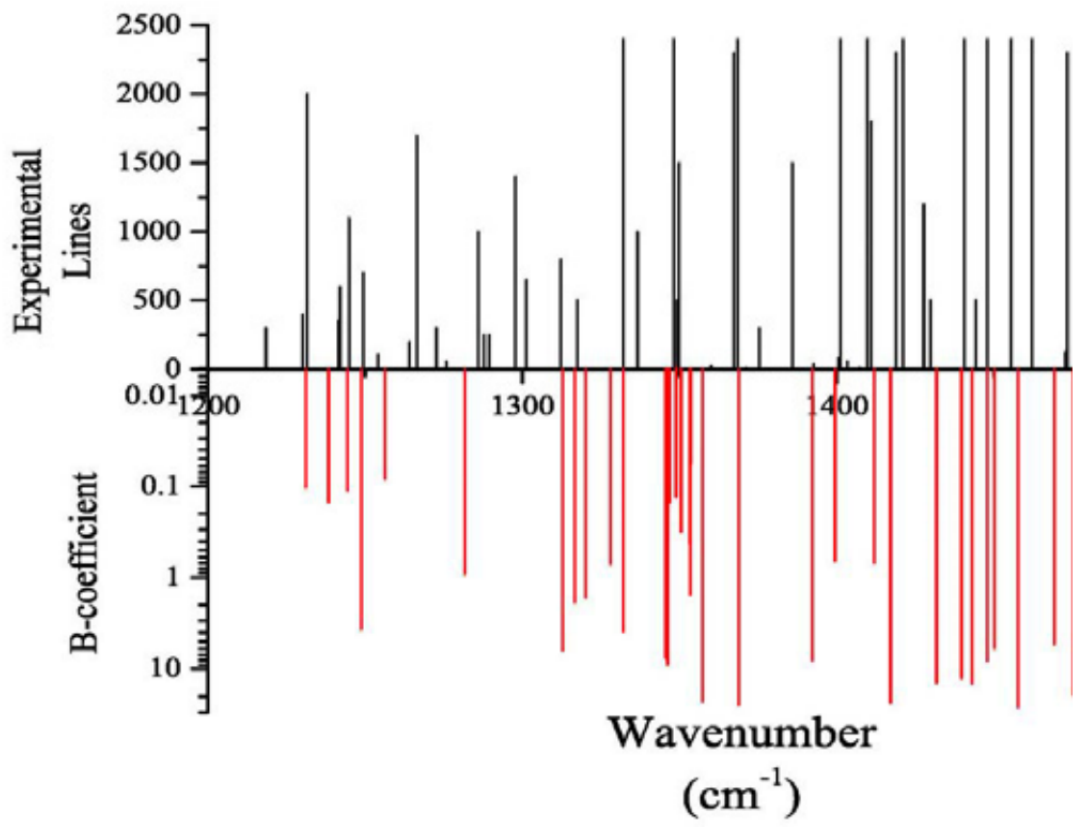

**Fig. 4:** Comparison of experimental and theoretical nd → np transitions of the $H_3$ Rydberg molecule. Intensities should not be compared directly. Experimental laser strength is in arbitrary units while theoretical stimulated emission B-coefficients are in the units of 1022 m $J^{-1}$ $S^{-2}$ on a logarithmic scale (Saykally et al., 2015).

In 2012, Galli and Palla summarized model cooling function calculations for various molecules. At low temperatures, the importance of $H_3^+$ becomes evident as an important coolant for primordial gas. While experimental data for actual $H_3$ Rydberg cooling remains elusive, the observed spectra and associated calculations for $H_3^+$ do suggest a potentially important role for $H_3$ Rydberg molecules in the cooling process.

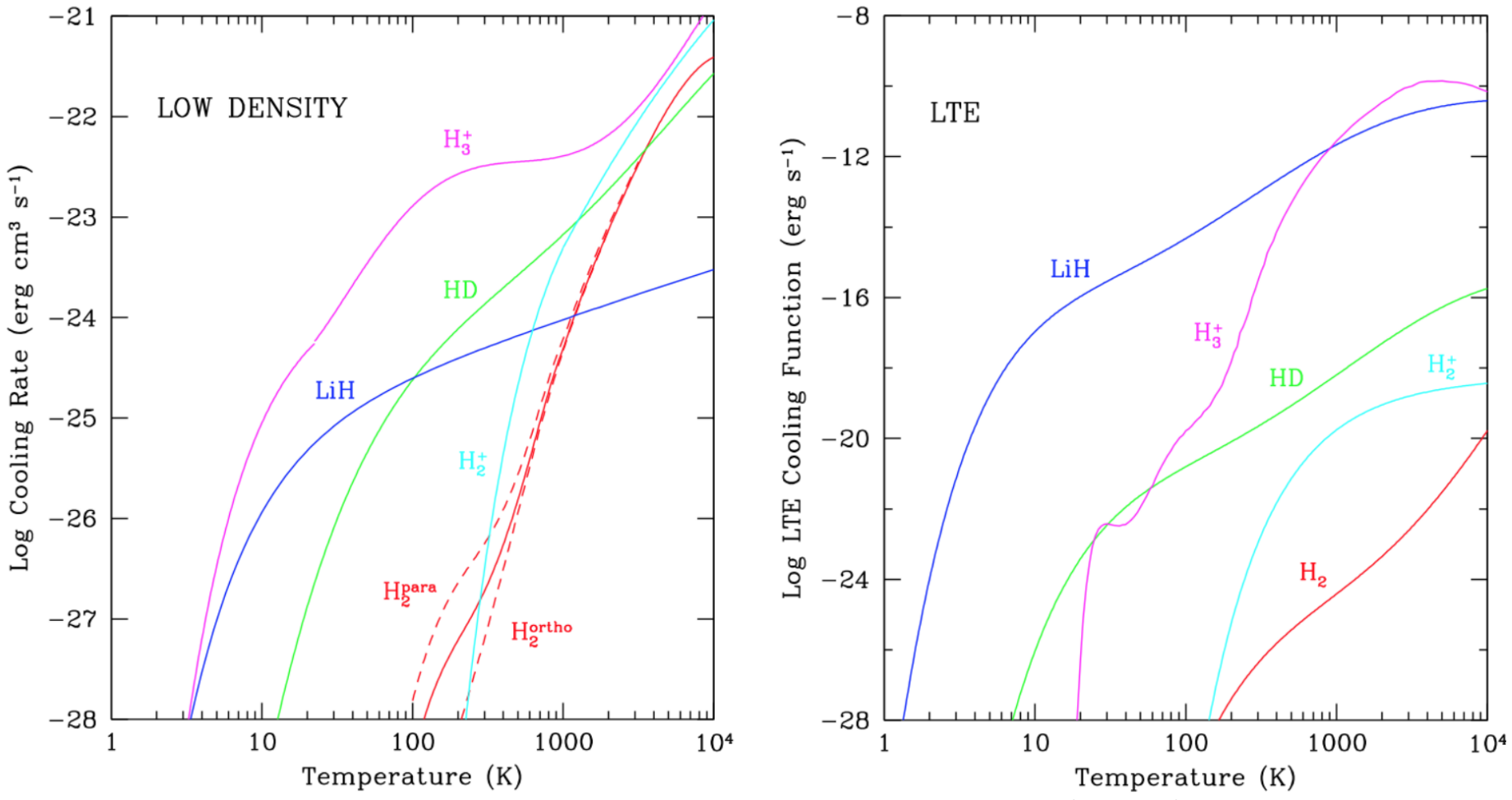

**Fig. 5:** Log cooling rate as a function of temperature for $H_2$, $H_2^+$, $H_3^+$, HD, and LiH. a) describes low-density limits, and b) describes the LTE limit. a) plots the cooling rate assuming an ortho-para ratio of 3:1. Sources for the individual low-density cooling rates are: Glover & Abel (2008) for $H_2$; Suchkov & Shchekinov (1978) and Glover & Savin (2009) for $H_2^+$; Glover & Savin (2009) for $H_3^+$; Lipovka, Núñez-López & Avila-Reese (2005) for HD; Bougleux & Galli (1998) for LiH.

Sources for the LTE cooling functions are: Coppola et al. (2012) for H2; Glover & Abel (2008) for $H_2^+$; Glover & Savin (2009) for $H_3^+$; Coppola et al. (2011) for HD and LiH. Figure from Galli & Palla (2012).

During 2011-2013, a series of papers published by Glosik, showed slightly increasing selectivity and other factors, comparing flowing and stationary afterglow plasmas. However, the lack of storage ring data below 300 K persisted (Petrignani, 2011). Using normal and para-enriched hydrogen, Glosik and coworkers were able to control the ratio of para- and ortho-$H_2$ to determine the binary recombination rates, and later, ternary recombination rates for both pure para- and ortho-$H_3^+$ ions. Population of $H_3^+$ in the para states is also temperature dependent (Varju et al., 2011) (Dohnal et al., 2012). The first recombination rate coefficients were determined for ions in the lowest ground vibrational state, with specific nuclear and rotational states measured *in situ*. These low-energy, state-specific recombination coefficient rates are in excellent agreement with theoretical predictions of binary DR. These efforts facilitated state-specific $H_3^+$ dissociative recombination experiments. At low temperatures (77-200 K), para- and ortho-$H_3^+$ dissociative recombination rates increasingly deviate as temperature decreases (Dohnal et al., 2012). Para-$H_3^+$ DR rates increase, while ortho-$H_3^+$ rates decrease. As such, the dissociative recombination patterns were experimentally determined below 100 K. Later, in 2015, it would be found that the 77 K binary recombination rate for para-$H_3^+$ ($1.5 \times 10^{-7}$ cm$^3$ s$^{-1}$) is five times larger than for ortho-$H_3^+$ ($3 \times 10^{-8}$ cm$^3$ s$^{-1}$).

In 2013, studies extended conditions down to 50 K for the first time for both $H_3^+$ and $D_3^+$ DR. Differing from neutral ternary recombination (He assisted), electron assisted ternary recombination (electron-collisional radiative recombination or E-CRR) may also substantially contribute to effective recombination rates, which was suggested by theory and observed in an $Ar^+$ dominated plasma. However, the contribution of E-CRR to the effective dissociative recombination rate could not be extracted in $H_3^+$ dominated plasma, even at 50 K (Johnsen et al., 2013). Moreover, E-CDR & N-CDR processes do not adequately explain existing experimental results. CRR at low temperatures competes with binary recombination. Collisional radiative recombination is a bigger factor than collisional dissociative recombination because CDR is only effective at low *l* states, while CRR occurs for both low and high *l* states and is statistically favored.

## The HeH Rydberg Molecule

Recently, the role of HeH+ as "the very first molecule to have been synthesized in the universe" has been elucidated via its discovery by radioastronomy (Güsten et al., 2019). HeH+ is generally considered important in the chemistry of the early universe, primarily as an intermediate in the synthesis of H2 (Lepp et al., 2002):

$$H^+ + He \rightarrow HeH^+ + \nu$$
$$HeH^+ + H \rightarrow H_2^+ + He$$
$$H_2^+ + H \rightarrow H_2 + H^+$$

Here, we investigate its role as a potential primordial gas coolant. With a large dipole and a potential recombinant Rydberg state manifold, $HeH^+$ could indeed prove to be an interesting target for such a study. At high redshift, $HeH^+$ exhibits a fractional abundance multiple orders of magnitude higher than that of $H_3^+$, but still 5-10 orders of magnitude lower than that of neutral $H_2$. Similar to the case of $H_3^+$, electron-$HeH^+$ recombination processes could lead to an excited molecular core, IR and/or UV-Vis radiation, and subsequent cooling. If neutral HeH Rydberg molecules do form, it may be necessary to account for such cooling effects.

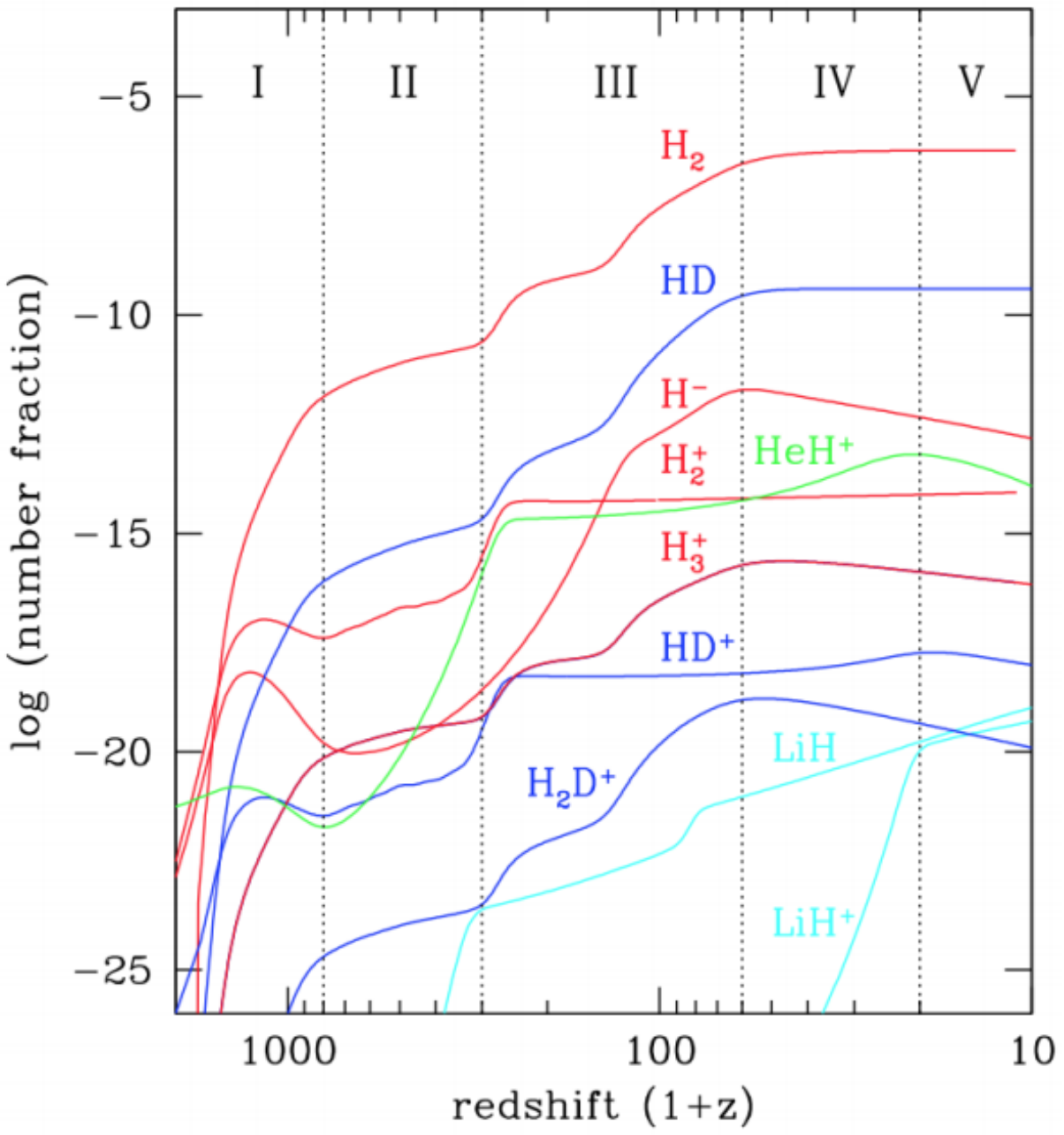

**Figure 6:** Fractional abundances of main species formed in the early universe as a function of redshift. Vertical lines indicate boundaries of the five main evolutionary phases, as described in Galli & Palla 2012 (Galli & Palla, 2012).

Recent (2016) *ab initio* theoretical studies have addressed such Rydberg states, and have shown that they can evolve regularly, similar to the Born-Oppenheimer function for $H_2$ (Bouhali et al., 2016). Herzberg, in 1977, was able to experimentally characterize the lower energy levels of $HeH^+$. This was possible due in part to the large dipole moment and its derivative, which generates strong rovibrational spectra in the infrared region. From these, it is possible to calculate the radiative energy loss from electron recombination (Dabrowski & Herzberg, 1977).

The cooling functions for $HeH^+$, $H_3^+$, and other such molecules have previously been described. To evaluate the cooling function as a function of redshift, knowledge of the radiative and collisional transition probabilities and the frequencies of rovibrational transitions of each molecule is necessary. These data have been reported in the literature for common molecules and ions. The rate equations can thus be solved for the energy level populations. However, accurate collisional coefficients are available in only a limited number of cases, and approximations must be used for minor species, such as $HeH^+$. The radiative cooling contribution for $HeH^+$ was described by Coppola, Lodi, and Tennyson in 2011, and is shown in Figure 7 (Coppola et al., 2011).

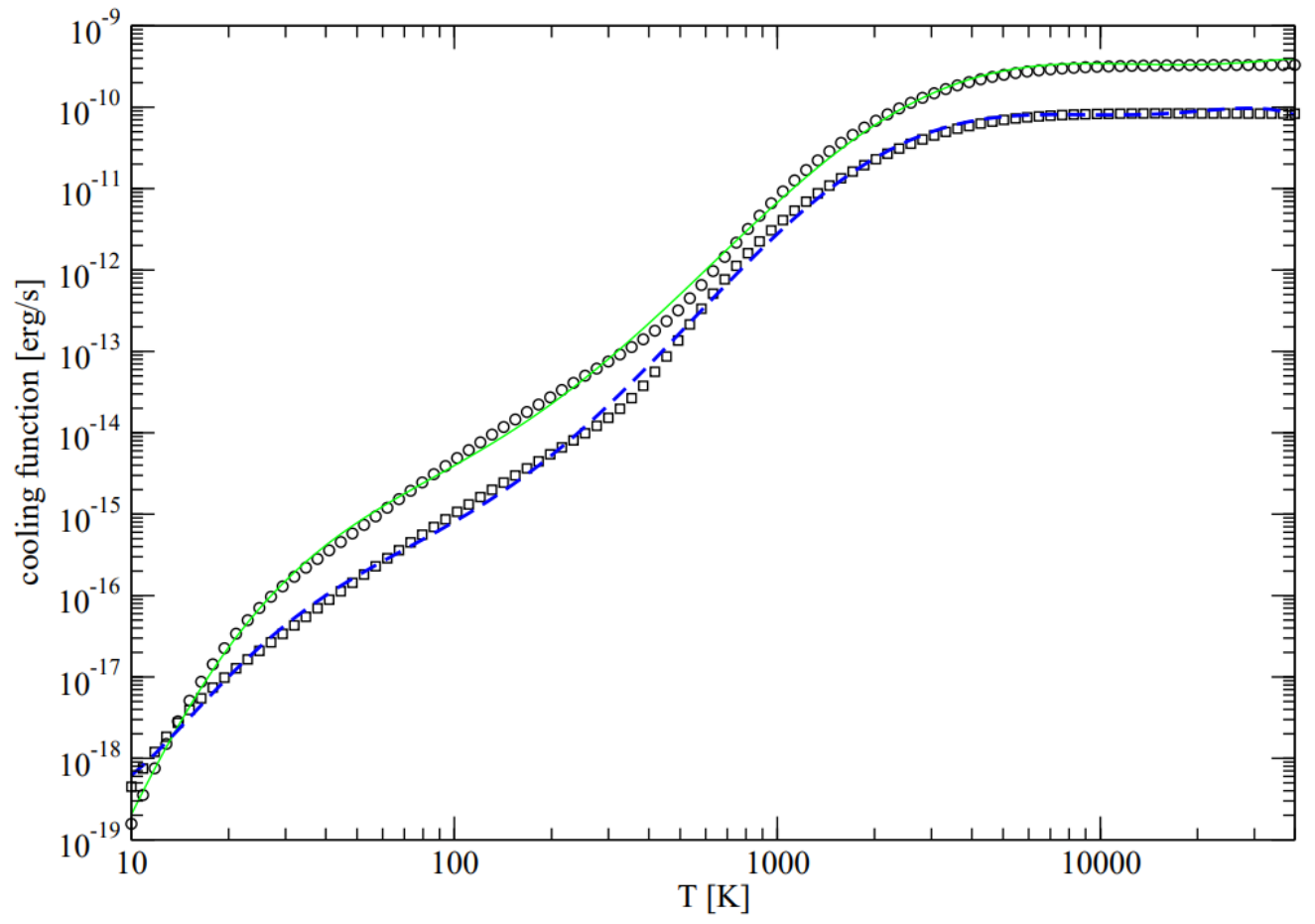

**Fig. 7:** $HeH^+$ radiative cooling function contributions. Circles: 3 $HeH^+$; squares: 3 $HeD^+$; green solid curve: 3 $HeH^+$ fit (Table 1); blue dashed curve: 3 $HeD^+$ fit (Coppola, Lodi, Tennyson 2011).

The cooling effects of $HeH^+$ ostensibly contribute non-negligibly to the overall cooling, likely attributable to its strong dipole moment. While HeH Rydberg molecules retain the strong dipole moment, they additionally have the ability to radiate large quantities of energy upon recombining with an electron, as they decay from high to low *n*-states. Hence, HeH Rydberg molecules comprise an interesting candidate for further investigation of potential coolants in primordial gas.

In order to obtain astronomical data to exploit these laboratory and theoretical observations, it would be interesting to examine the 21cm$^{-1}$ hydrogen line to explore conditions at relevant redshifts, and then to ultimately scan appropriate frequency regions for emissions from candidate primordial molecules. In 1944, Hendrik van de Hulst predicted that atomic hydrogen should emit (van Woerden, H. & Strom, R.G., 2006) at a wavelength of 21.1 cm$^{-1}$ corresponding to a transition between hyperfine levels of the 1s ground state in a spin-flip transition. While this magnetic dipole transition is very weak, the high abundance of neutral hydrogen atoms makes it possible to detect it using modern radio telescopes. Very interestingly, highly redshifted (z ~17) 21 cm$^{-1}$ emission was recently detected at 78 MHz by the radiofrequency array in Australia (Bowman et al., 2018). This light probes the universe only 180 million years after the Big Bang, as stars would have formed and emitted a background of Lyman-alpha radiation, giving rise to the detected signal. Until now, it was believed that adiabatic cooling due to expansion of the universe was the primary initial cooling mechanism. While the observed line profile is consistent with what is expected of population III stars, the amplitude of the absorption is 0.5 K, which is more than twice the predicted value due to adiabatic cooling (the excess significance is 3.8σ per Barkana, 2018). Another possibility is a much hotter background, but this is unlikely since precise measurements of the background radiation from Planck Collaboration et al. 2019 indicate temperature anisotropies in line with estimates. Bowman and coworkers, in 2018, reference Barkana (2016) who invokes velocity-dependent Rutherford/Coulomb scattering interactions between baryons and cold dark matter to explain the cooling and the large drop in amplitude of the absorption profile. However, Barkana and coworkers (2018) negated the possibility due to the requirement of strong interactions at z~20 and dominant dark matter to be cooled enough to explain the signal anomaly, pointing to the existence of additional cooling mechanisms.

If $H_3$ and HeH Rydberg molecules did indeed produce strong emissions over large pathlengths from such massive objects as primordial stars, the resultant light may currently be detectable. Using the 21 $cm^{-1}$ hydrogen line as a guide, we can estimate, shifting the laboratory 7 micron emission from $H_3$ into the far IR, and the 0.2-0.4 micron HeH emission into the mid-IR.. Examination of these regions with appropriate telescopes such as SOFIA and other Hi-Fi instruments could yield interesting new insights into proposed new cooling mechanisms...

## Conclusions

While $H_2$ associative detachment remains the accepted primary cooling mechanism for star formation in primordial gas, the weakness of quadrupole-allowed $H_2$ rotational transitions remains a concern regarding cooling to the requisite low temperatures required for collapse to fusion densities. In the search for alternative/additional coolants, $H_3^+$ and $HeH^+$ have been discussed. The laboratory observation and characterization of neutral $H_3$ Rydberg lasing processes in supersonic hydrogen/helium plasmas reveals another candidate for cooling. Rydberg molecules can radiate large amounts of energy on relatively short timescales. As such, the cooling per molecule could be extremely high, and Rydberg molecules could thus play an important role-even at their relatively low abundances. In addition to the $H_3$ Rydberg species, an increased understanding of HeH in recent years presents another interesting candidate for cooling. With a strong dipole moment and derivative, as well as a Rydberg architecture, the HeH Rydberg molecule could effect for high molecular cooling rates at relatively low abundances. We anticipate further experimentation into such Rydberg molecules in order to characterize their radiative cooling profiles, as well as further theoretical investigation into cooling functions of such molecules to elucidate their potential influence in primordial gas clouds.

Knowing that $H_3^+$ is formed in significant densities, we can infer that $H_3^+$ recombination with electrons to produce $H_3$ Rydberg molecules may have been prominent in the early universe. Strong IR emissions from these molecules would have occurred and be detectable at very high redshifts. The IR lasing results reported by Saykally et al. would be shifted deep into far-IR/microwave region of the spectrum. High resolution telescope imaging of the 200-500 micron region for $H_3$ Rydberg and the 0.1-0.4 micron region for HeH Rydberg could yield interesting results.

By searching for transitions of these Rydberg molecules in the context of primordial cooling, we could advance our understanding of the formation of the first stars in the universe. In order to assess the cooling potential of these novel systems, detailed chemical reaction modeling must be performed, and similarly detailed radiative transfer models must be developed. We hope that this brief review helps to stimulate such efforts.

## Acknowledgments

We warmly thank Dr. Tim Lee (NASA-Ames ), Prof. Xander Tielens (Leiden University) and Dr. Paola Caselli (Max Planck Institute for Extraterrestrial Physics) for their insightful comments on this manuscript.